\documentclass[aps,prb,nofootinbib,superscriptaddress,showpacs,reprint,floatfix]{revtex4-2}
\usepackage{amsmath}
\usepackage{physics}
\usepackage{amssymb}
\usepackage[inline]{trackchanges}
\allowdisplaybreaks[4]
\pagebreak
\usepackage{amsfonts}
\usepackage[usenames,dvipsnames]{pstricks}
\usepackage{mathtools}
\usepackage[colorlinks,hyperindex]{hyperref}
\usepackage{array}

\newcommand{\bk}{\mathbf{k}}
\newcommand{\bG}{\boldsymbol{\Gamma}}
\newcommand{\bM}{\mathbf{M}}
\newcommand{\ketZ}{\ket{\mathbb{Z}_2}}
\newcommand{\omegaS}{\omega_\text{scar}}
\newcommand{\nuS}{\nu_\text{scar}}
\newcommand{\Hfull}{H_\text{full}}
\newcommand{\br}{\mathbf{r}}

\hypersetup
{	colorlinks,%
	citecolor=green,%
	linkcolor=red,%
	urlcolor=blue%
}

\newcommand{\su}{\mathfrak{su}}

\begin{document}
\title{Stability of scar states in 2D PXP model against random disorders}
\author{Ke Huang}
\affiliation{Department of Physics, City University of Hong Kong, Kowloon, Hong Kong SAR}
\affiliation{School of Physics and Technology, Wuhan University, Wuhan 430072, China}
\author{Yu Wang}
\affiliation{School of Physics and Technology, Wuhan University, Wuhan 430072, China}
\author{Xiao Li}
\email{xiao.li@cityu.edu.hk}
\affiliation{Department of Physics, City University of Hong Kong, Kowloon, Hong Kong SAR}
\affiliation{City University of Hong Kong Shenzhen Research Institute, Shenzhen 518057, Guangdong, China}

\date{\today}

\begin{abstract}
Recently a class of quantum systems exhibiting weak ergodicity breaking has attracted much attention. 
These systems feature in the energy spectrum a special band of eigenstates called quantum many-body scar states. 
One prototype model that is known to host scar states is the so-called PXP model. 
In this work we study the fate of quantum many-body scar states in a 2D PXP model against random disorders. 
We show that in both the square lattice and the honeycomb lattice the scar states can persist up to a finite disorder strength, before eventually being erased by the disorder. 
We further study the localization properties of the system in the presence of even stronger disorders and show that whether a full localization transition occurs depends on the type of disorder we introduce. 
Our study thus reveals the fascinating interplay between disorder and quantum many-body scarring in a 2D system. 
\end{abstract}

\maketitle

\section{Introduction}
Recently rapid experimental progress in creating and manipulating isolated quantum systems~\cite{Bloch2008_Review,Georgescu2014_Review} has enabled the simulation of novel quantum phases of matter in out-of-equilibrium systems.
In particular, whether an isolated quantum system can reach thermal equilibrium under its own dynamics is currently one of the central questions in condensed matter physics. 
The answer to this question is crucial because it is closely related to the ergodicity hypothesis, which is one of the fundamental postulates of statistical mechanics. 
To date the eigenstate thermalization hypothesis (ETH)~\cite{Deutsch1991,Srednicki1994,Rigol2008} provides the most plausible mechanism for thermalization in an isolated quantum system. 
In particular, it states that thermalization occurs at the eigenstate level: The expectation value of local observables in each eigenstate is thermal.

Meanwhile, several systems that exhibit ergodicity breaking have been identified, including noninteracting systems and certain fine-tuned integrable models~\cite{Kinoshita2006}. 
{Furthermore, it has been realized that strong disorders can provide a generic and robust mechanism for ergodicity breaking.} 
In particular, the localization can persist even when finite interactions are present, leading to the so-called many-body localization (MBL) phase~\cite{Nandkishore2015_Review,Altman2015_Review,Abanin2019_Review,Gopalakrishnan2020_Review}.  
One hallmark of the MBL phase is that ergodicity breaking occurs for all typical eigenstates. 
In particular, it is believed that in an MBL system quench dynamics from a generic initial nonequilibrium state is nonergodic, and slow relaxations can be expected. 

In the past few years, yet another form of ergodicity breaking has been identified, and rapid progress has been made to understand this phenomenon. 
In particular, {a recent experiment performed in an array of strongly interacting Rydberg atoms measured the return probability of various different initial states when nonequilibrium dynamics is initiated}~\cite{Bernien2017_Nature}. 
It was found that persistent revivals occur when the quench dynamics starts from a specific initial state---the $\ketZ$ state ({defined as the state where all atoms on even lattice sites are in the excited state, while those on odd lattice sites remain in their ground states}), whereas a much faster relaxation is observed in the quench dynamics from most other initial states. 
The fact that relaxation {in} this system has a strong dependence on the initial conditions indicates that this system is qualitatively different from an MBL system. 
It turns out that the persistent revival can be attributed to a special band of eigenstates in the energy spectrum that are {nonthermal} (characterized by their low entanglement entropies), 
which are now dubbed as \emph{quantum many-body scar states}~\cite{Turner2018_NatPhys,Turner2018_PRB,Lin2020_PRB,Michailidis2020_PRR,Moudgalya2018_PRB,Ho2019_PRL,Michailidis2020_PRX,Iadecola2019_PRB,Choi2019_PRL,Khemani2019_PRB,Lin2019_PRL,Bull2019_PRL,Ok2019_PRR,Bull2020_PRB,Zhao2020_PRL,Mukherjee2020_PRB,McClarty2020_PRB,Srivatsa2020_PRB,Kuno2020_PRB,Lin2020_PRR,Moudgalya2020_PRB,Pakrouski2020_PRL,Surace2021_PRB,Banerjee2021_PRL,Wildeboer2021_PRB,Serbyn2021_NatPhys,MondragonShem2021,Zhao2021_PRL}.  
Several interesting properties of scar states have been uncovered so far. 
For example, it was shown that the scar states are associated with a hidden unstable periodic orbit~\cite{Ho2019_PRL,Michailidis2020_PRX}, in analogy to a classically chaotic system~\cite{Heller1984_PRL}. 
In addition, an approximate $\su(2)$-algebra has been identified in the PXP model~\cite{Turner2018_NatPhys,Turner2018_PRB,Lin2020_PRB,Michailidis2020_PRR}, which represents a good approximation for the scar states in the original Rydberg Hamiltonian. 
Subsequently, similar ``spectrum generating algebras''~\cite{Serbyn2021_NatPhys} have been identified in a variety of other models, such as the Afflect-Kennedy-Lieb-Tasaki (AKLT) model~\cite{Moudgalya2018_PRB}, and several other noninteracting spin models~\cite{Iadecola2019_PRL,Schecter2019_PRL,Iadecola2020_PRB,Chattopadhyay2020_PRB,Shibata2020_PRL}.

One of the central open questions in this field is the stability of these scar states against various perturbations~\cite{Turner2018_PRB,Lin2020_PRR,Surace2021_PRB,Moudgalya2020_PRB,Pakrouski2020_PRL,MondragonShem2021}, and much effort has been devoted to this question. 
{For example, it was shown that the scar states remain stable if the perturbations are compatible with the so-called forward scattering approximation~\cite{Turner2018_PRB}. 
In addition, the interplay between 2D scar states and disorders in Hubbard-like models has been studied~\cite{Moudgalya2020_PRB,Pakrouski2020_PRL}. 
Moreover, the response of the scar states in 1D PXP model to generic perturbations was analyzed by studying how the fidelity scales with the size of the system~\cite{Surace2021_PRB}.} 
However, the question of whether scar states in the PXP model are stable against random external disorders has not been sufficiently addressed in the literature. 
One notable recent work along this direction carefully analyzed the stability of scar states in the 1D PXP model against random disorders~\cite{MondragonShem2021}, and it was shown that the scar states are stable against finite disorders. 
Yet, the stability of scar states in the 2D PXP model~\cite{Lin2020_PRB,Michailidis2020_PRR} in the presence of random disorders has not been addressed in the literature.  
This question is crucial for two reasons. 
First, the effect of random disorders in a 1D system is often qualitatively different from that in higher dimensional systems. 
For example, it is generally believed that MBL is a stable phase in a 1D disordered system~\cite{Imbrie2016}, whereas its existence in a 2D system is still an open question. 
Second, while the existence of 2D scar states have been predicted in several pioneering works~\cite{Lin2020_PRB,Michailidis2020_PRR}, its experimental observation is still elusive. 
In this regard, it is important to understand the stability of such 2D scar states against various perturbations, especially the random disorders, which are ubiquitous in the experiments. 
Therefore, it is important to analyze this question carefully. 

In this work, we study the robustness of revivals from a $\ketZ$ state in a 2D PXP model against random disorders. 
We consider both the square lattices and the honeycomb lattices. 
We find that scar states in the 2D PXP model are generically stable against finite random disorders before it is eventually erased at stronger disorders. 
We also analyze the fate of the PXP model itself against very strong disorders and find that whether the system eventually enters a many-body localized phase depends on the type of disorder present in the system. 

\section{Model} 
To begin with, we introduce the Hamiltonian for Rydberg atoms in a 2D square lattice of $N$ lattice sites, which can be written as follows~\cite{Turner2018_NatPhys,Turner2018_PRB}
\begin{align}
    H_R=\sum_{\br}\qty\bigg[\dfrac{\Omega}{2}\sigma_{\br}^x+J_RP_{\br}^+\sum_{\langle\br',\br\rangle}P_{\br'}^{+}].  
\end{align}
In the above equation, $\br=(i,j)$ goes over all lattice sites and $\langle\br',\br\rangle$ denotes summation over nearest neighbors of $\br$. 
In addition, $\sigma_{\br}^x=\ketbra{\bullet}{\circ}+\ketbra{\circ}{\bullet}$ represents a Pauli matrix describing the transition between the ground state $|\circ\rangle$ and the excited state $|\bullet\rangle$ at site $\br$. 
The projection operators $P_{\br}^{+}=\ketbra{\bullet}{\bullet}$ and $P_{\br}^{-}=\ketbra{\circ}{\circ}$ project states into their excited or ground parts respectively. 
Finally, the Rabi frequency $\Omega$ represents the rapidity of the oscillations between the excited state and the ground state, and $J_R$ exerts an extra energy if two adjacent sites are both excited. 
Throughout this work we will maintain periodic boundary conditions in both directions.

In the limit $J_R\gg \Omega$, the probability that two nearest neighbors are both excited is strongly suppressed due to the significant energy cost. 
Hence, in this strong interaction limit, the low-energy states in the system are almost limited to the subspace spanned by states without any nearest neighbors simultaneously in the exited state. 
As a result, the system can be captured by the so-called PXP model~\cite{Turner2018_NatPhys,Turner2018_PRB,Lin2020_PRB,Michailidis2020_PRR}
\begin{align}
    H_{\mathrm{PXP}}=\mathcal{P}H_R\mathcal{P},
\end{align}
where $\mathcal{P}=\prod_{\langle\br',\br\rangle}(\mathbb{I}-P_{\br}^+P_{\br'}^+)$. Equivalently, this Hamiltonian can be expressed as
\begin{align}
    H_{\mathrm{PXP}}=\frac{\Omega}{2}\sum_{\br}\sigma_{\br}^x\prod_{\langle\br',\br\rangle}P_{\br'}^-.
\end{align}
The above Hamiltonian possesses a separated band of $N+1$ eigenstates known as {quantum many-body scar states}, and they have high overlaps with $\ketZ$, which is the maximally excited state in this subspace. 
Therefore, the quench dynamics starting from $\ketZ$ is mostly controlled by these $N+1$ scar states. 
Furthermore, because of an $\su(2)$-like algebraic structure in $H_{\mathrm{PXP}}$~\cite{Choi2019_PRL,Khemani2019_PRB,Bull2020_PRB}, the energies of the scar states are separated almost evenly, resulting in a strong revival of $\ketZ$.

However, the $\su(2)$ algebra in the pristine PXP model is only approximate and the energies of scar states are not perfectly aligned, especially in a 2D lattice. 
It has been pointed out that one can enhance the $\su(2)$-like structure and stabilize the revival of $\ketZ$ in a square lattice by adding some weak perturbations to the PXP model~\cite{Michailidis2020_PRR}
\begin{align}
    \delta H=\sum_{\br}\sigma_{\br}^x\left[aP_{\br}^l+2aP_{\br}^d + bP_{\br}^3\right]\prod_{\langle\br',\br\rangle}P_{\br'}^-,
\end{align}
where $a\approx0.012\Omega$ and $b\approx0.027\Omega$.
{In addition, $\br \equiv (i,j)$ is a 2D vector that labels different sites on the 2D lattice.} 
Moreover, the various operators in the above equations are given by~\cite{Michailidis2020_PRR},
\begin{align}
    P_{i,j}^l &= P_{i+2,j}^-+\cdots, \quad 
    P_{i,j}^d = P_{i+1,j+1}^-+\cdots,\\
    P_{i,j}^3 &= P_{i+1,j+1}^-P_{i+1,j-1}^-P_{i+2,j}^-+\cdots.\notag
\end{align}
{Note that here we have explicitly written down the components of each vector $\br = (i,j)$. 
In addition, for each equation, the ellipses denote the other three terms derived by consecutive $\pi/2$ rotations, which preserve the $C_4$ rotational symmetry of the 2D square lattice. 
The total Hamiltonian for a clean PXP model thus reads
}
\begin{align}
    H_\text{clean} = H_\text{PXP} + \delta{H}. \label{Eq:CleanModel}
\end{align}

In order to study the stability of the scar states, especially the periodic revivals, against random disorders we introduce the following disorder perturbation to the Rydberg Hamiltonian $H_R$, 
\begin{align}
    H_w=\sum_{\br}h_a(\br)\sigma_{\br}^a, \label{Eq:Disorder_Full}
\end{align}
where $a=x,y,z$ and all components of $h_a(\br)$ are uniformly distributed in $[-W/2,W/2]$. 
{
This particular disorder is chosen because it is one of the most general perturbation we can write down at the single-particle level. 
In particular, it explicitly breaks the symmetries of the pristine PXP model such as time-reversal symmetry. 
}
Because we focus on the strong interaction limit in this work, we will also apply to the disorder potential the rule that no two neighboring sites can be simultaneously excited. 
As a result, we can write the full effective Hamiltonian as follows, 
\begin{align}
    \Hfull=H_{\mathrm{PXP}}+\delta H+\mathcal{P}H_w\mathcal{P}, \label{Eq:FullModel}
\end{align}
which is the main model we study in this work. 
{In Sec.~\ref{Section:MBL} we will further discuss the properties of the system when one or more of $h_a(\br)$ is zero in order to present a more complete picture.} 

\begin{figure}[t]
\includegraphics[width=\columnwidth]{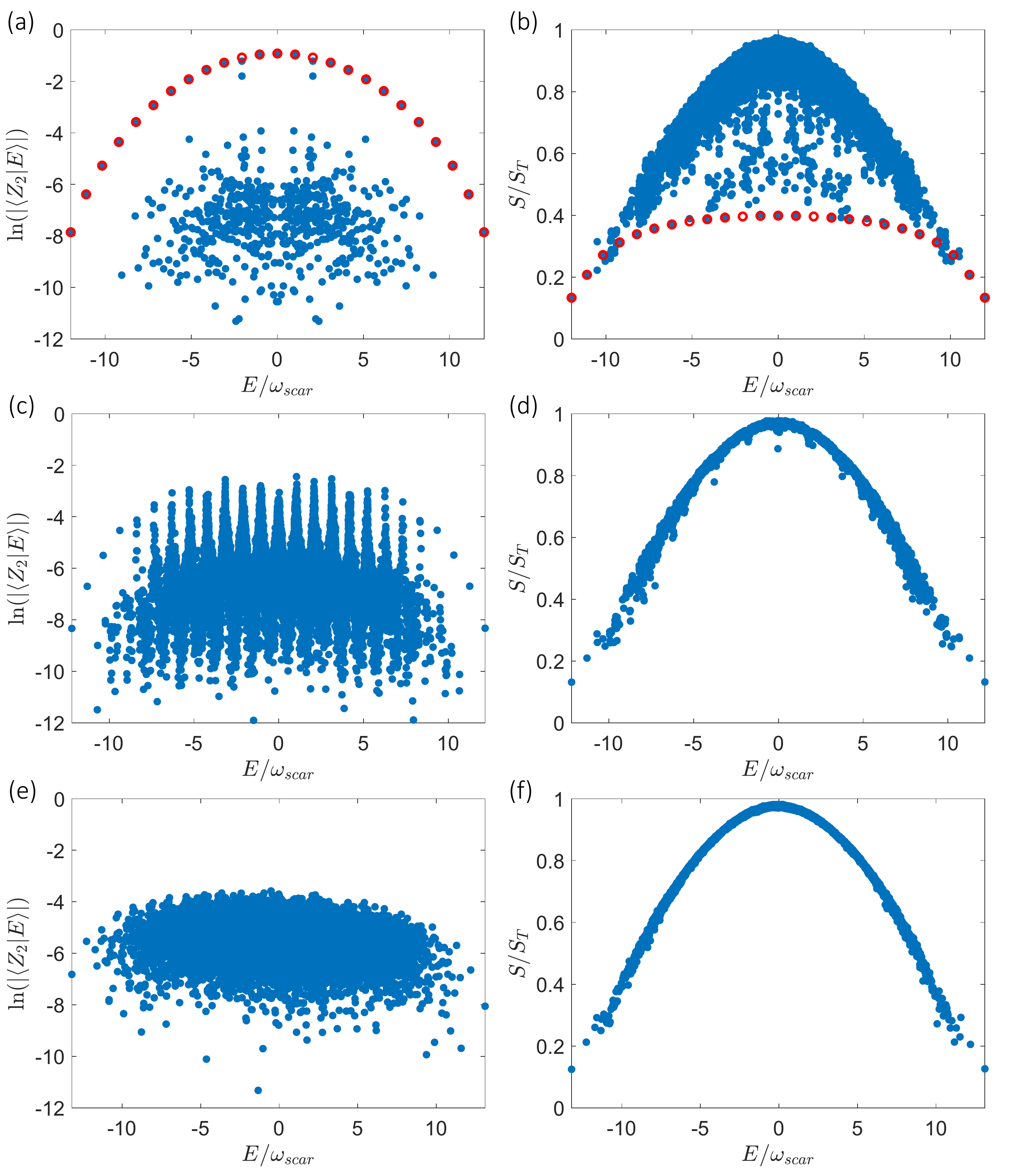}
\caption{\label{Fig:SquareLattice} Left panel: the $\ketZ$ state in a $6\times4$ square lattice. 
Right panel: The EE of all eigenstates in the same model as the corresponding left plot, where $S_T$ is the entanglement Page value (see text). 
From top to bottom, the two panels in each row represent the system with a fixed disorder realization with $W=0,0.1\omegaS,0.5\omegaS$, respectively. 
Finally, blue dots are obtained by exact diagonalization methods and red circles in (a), (b) are results from the FSA approximation.}
\end{figure}

\section{Stability of the scar states}

\subsection{Spectral properties}

One hallmark of scar states in {the 1D PXP model} is that $\ketZ$---{the special initial state that gives rise to persistent revivals in the experiments}---approximately resides in the subspace spanned by the scar states. 
Here we show the existence of quantum many-body scar states in the model we consider [see Eq.~\eqref{Eq:FullModel}] and demonstrate that the scars in the square lattice share similar properties to those in the 1D PXP model. 
{In particular, we show that $\ketZ$ again approximately resides in the subspace spanned by the scar states and that persistent revivals from the $\ketZ$ can also occur in this model}. 

In Fig.~\ref{Fig:SquareLattice}(a), we show that in a clean $6\times4$ square lattice [with the Hamiltonian given by Eq.~\eqref{Eq:CleanModel}], a band of eigenstates can also be identified as scar states due to their outstanding overlap with $\ketZ$. 
More importantly, the energies of these scar states are spaced evenly by a frequency $\omegaS=2\pi\nuS=0.6773\Omega$, which results in an almost periodic revival.
The existence of scar states are further affirmed by their low {entanglement} entropy (EE), as shown in Fig.~\ref{Fig:SquareLattice}(b). 
Specifically, we divide the $6\times4$ square lattice into a left half $L$ ($i\leq3$) and a right half $R$ ($i\geq4$).
Under this partition, we can calculate the entanglement entropy of each eigenstate $S(|\phi\rangle)=-\mathrm{Tr}\left\{\rho_L(|\phi\rangle)\ln\rho_L(|\phi\rangle)\right\}$, where $\rho_L(|\phi\rangle)=\mathrm{Tr}_R(|\phi\rangle\langle\phi|)$. 
In Fig.~\ref{Fig:SquareLattice}(b), the EE of these scar states are clearly much smaller than that of the rest of the eigenstates. 
{Therefore, we can see that the scar states in the clean PXP model have three important features: (1) strong overlaps with $\ketZ$, (2) evenly spaced energies, and (3) low entanglement entropoy. }

We have also verified that in the clean limit the forward scattering approximation (FSA)~\cite{Turner2018_NatPhys,Turner2018_PRB,Choi2019_PRL,Michailidis2020_PRR} remains applicable for both the overlap and EE in this 2D square lattice. 
{
This approximation is very useful for understanding the physics behind scar states. 
Essentially, it is a modified form of the Lanczos iteration procedure. 
It allows one to start from the state $\ketZ$ and iteratively generate $N+1$ special states, where $N$ is the number of lattice sites in the system. 
These $N+1$ states span a subspace $\mathcal{K}$, in which the scar states live predominantly~\cite{Turner2018_NatPhys,Turner2018_PRB}. 
}

\begin{figure*}[t]
    \includegraphics[width=\textwidth]{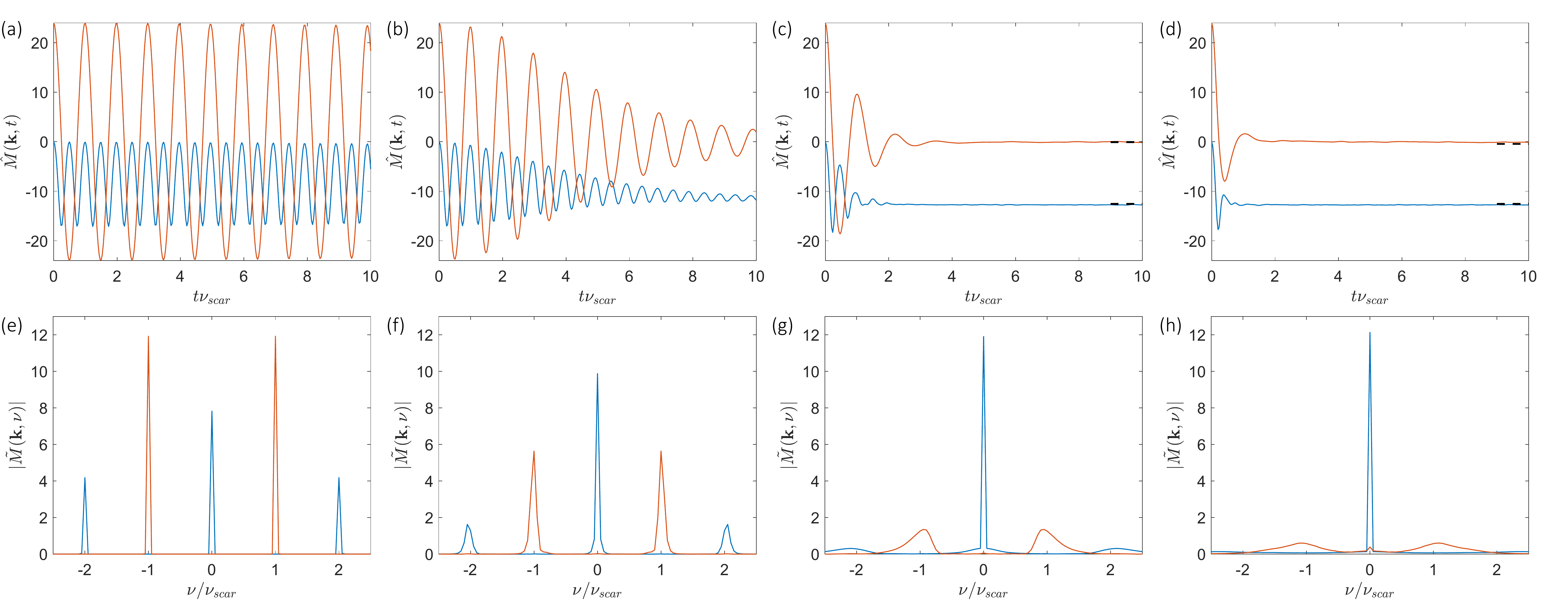}
    \caption{\label{Fig:Magnetization} 
    (a)-(d) show $\hat{M}(\textbf{k},t)$ in systems with a disorder strength of $W=0$, $0.1\omegaS$, $0.5\omegaS$, and $\omegaS$, respectively.
    Only one fixed disorder realization is used in generating these results. 
    (e)-(g) in the lower panel show the Fourier transformation of the corresponding upper panels. 
    The blue lines represent $\hat{M}(\textbf{k},t)$ or $\tilde{M}(\textbf{k},\nu)$ at $\boldsymbol{\Gamma}$, while the orange lines represent $\hat{M}(\textbf{k},t)$ or $\tilde{M}(\textbf{k},\nu)$ at $\mathbf{M}$.
    The canonical (microcanonical) values of $\hat{M}(\boldsymbol{\Gamma},t)$ (blue lines) in (c) and (d) are $-12.806$ ($-12.560$) and $-12.795$ ($-12.570$), respectively. 
    The canonical (microcanonical) values of $\hat{M}(\mathbf{M},t)$ (orange lines) in (c) and (d) are $0.005$ ($-0.040$) and $0.0207$ ($-0.435$), respectively. 
    Finally, the microcanonical values of $\hat{M}(\textbf{k},t)$ are marked by black dashed lines in (c) and (d).
    }
\end{figure*}

We now study the stability of these scar states against random disorders. 
When we turn on a weak disorder [$W=0.1\omegaS$, see Fig.~\ref{Fig:SquareLattice}(c)], more (but not all) eigenstates have an appreciable overlap with $\ketZ$, implying that the dynamics of $\ketZ$ is no longer controlled by just a few states. 
Actually, each scar state becomes a band of states and the energy of each scar state is broadened. 
Hence the evenly spaced spectrum still exist, suggesting a still oscillating but slowly decaying dynamics of $\ketZ$. 
However, though the spectrum resists the disorder, there are no more eigenstates with low EE in the spectrum now, as shown in Fig.~\ref{Fig:SquareLattice}(d). 
Instead, the majority of the eigenstates follows the volume law, approaching the Page value $S_T=\ln(181)-1/2$~\cite{Khemani2017_PRL} in a $6\times4$ square lattice. 
{To summarize, in the presence of a modest disorder, certain signatures of scar states in the clean system still remain, including (1) strong overlaps with $\ketZ$ and (2) evenly spaced energies. 
However, the low-entanglement feature is now lost. 
}

When we apply an even stronger disorder with $W=0.5\omegaS$ [see Fig.~\ref{Fig:SquareLattice}(e)], we find that most eigenstates now have a similar overlap with $\ketZ$, and that there is no special band of states with an evenly spaced energies. 
This observation implies that $\ketZ$ may lose its revival if the disorder strength is strong enough, as we verify in the following. 
Additionally, the EE in Fig.~\ref{Fig:SquareLattice}(f) obeys volume law, which is similar to Fig.~\ref{Fig:SquareLattice}(d). 
In particular, the EE now seems to be a continuous function of energy, which is reminiscent of the case of a thermal system.

\subsection{Quench dynamics from the \texorpdfstring{$\ketZ$}{Z2} state}

Another typical result of the scar states is the periodic revivals of the quench dynamics from $\ketZ$.
A useful quantity to characterize the dynamics of $\ketZ$ is the magnetization $M(\br,t)=\expval{\sigma_{\br}^z}{\mathbb{Z}_2(t)}$~\cite{MondragonShem2021}.
In particular, it is helpful to consider this quantity in its momentum space $\hat{M}(\textbf{k},t)$ as well as the corresponding Fourier transformation $\tilde M(\textbf{k},\omega)$, 
\begin{align}
    \hat{M}(\textbf{k},t)&=\sum_{\br}e^{i\textbf{k}\cdot\br}M(\br,t),\\
    \tilde M(\textbf{k},\omega)&=\int_{-\infty}^{+\infty}M(\textbf{k},t)e^{-i\omega t}\mbox dt.
\end{align}
Note that we sometimes use $\nu = \omega/(2\pi)$ instead of $\omega$. 

We first study the properties of $\hat{M}(\textbf{k},t)$ at two special momentum points $\bG=(0,0)$ and $\bM=(\pi,\pi)$. 
As shown in Fig.~\ref{Fig:Magnetization}(a), in a clean system the magnetization $\hat{M}(\textbf{k},t)$ at these two points oscillates {at a frequency of} $2\nuS$ and $\nuS$, respectively, and the oscillations show no sign of decay.  
This result is further confirmed by its Fourier transformation, where two peaks emerge at $\pm2\nuS$ and $\pm\nuS$, as shown in Fig.~\ref{Fig:Magnetization}(e). 

When a weak disorder is turned on [$W=0.1\omegaS$, Fig.~\ref{Fig:Magnetization}(b)], $\hat{M}(\textbf{k},t)$ still oscillates at the same frequency for many periods, but eventually decays, corresponding to the widening of the peaks in Fig.~\ref{Fig:Magnetization}(f). 
Moreover, we notice that the magnetization at $\bM$ is more resistant to disorder than its counterpart at $\bG$, as evidenced by its slower decay and stronger peaks of its Fourier transformation. 
{Thus, we have shown that the periodic revivals from $\ketZ$ can survive finite random disorders, which is consistent with the observation in Fig.~\ref{Fig:SquareLattice}(c) that certain features of the scar states still remain visible.} 

If the disorder is strong enough [$W\geq0.5\omegaS$, Fig.~\ref{Fig:Magnetization}(c) and~\ref{Fig:Magnetization}(d)] the magnetization loses all its nonergodic features, and quickly reaches thermal equilibrium. 
Correspondingly, $\tilde{M}(\bk,\nu)$ peaks at $\nu=0$ with a strength equal to their equilibrium values, as shown in Fig.~\ref{Fig:Magnetization}(g) and~\ref{Fig:Magnetization}(h), and only small bumps are left at $\pm\nuS$.

\begin{figure}[t]
\includegraphics[width=\columnwidth]{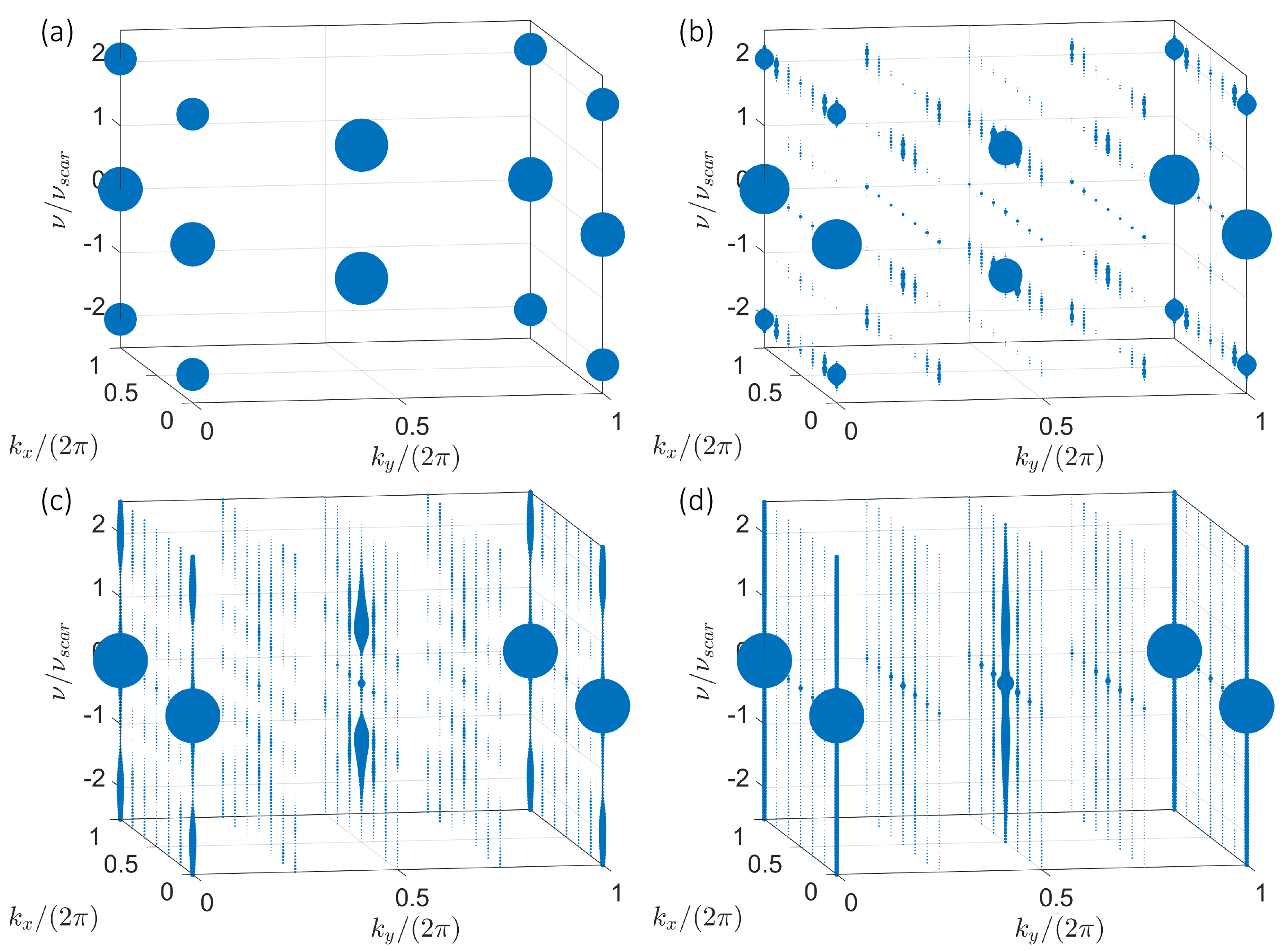}
\caption{\label{Fig:FourierAnalysis} 
(a)-(d) show $|\tilde M(\textbf{k},\omega)|$ in systems with $W/\omega_\text{scar}=0$, $0.1$, $0.5$, $1.0$, respectively. All results are averaged over $20$ disorder realizations.
The radius of each sphere is proportional to $\abs*{\tilde M(\textbf{k},\omega)}^{1/2}$ at each $\textbf{k}$ point. }
\end{figure}

We further consider the behavior of $\tilde M(\textbf{k},\omega)$ at generic $\textbf{k}$ points in the momentum space, in order to better understand the stability of the system and the effects of disorder. 
In Fig.~\ref{Fig:FourierAnalysis} we calculated $\tilde M(\bk,\omega)$ for four different disorder strengths $W=0,\,0.1\omegaS,\,0.5\omegaS,\,\omegaS$. 
The results are all averaged over $100$ disorder realizations when $W\neq 0$.  
We first look at the $W=0$ limit in Fig.~\ref{Fig:FourierAnalysis}(a). 
Due to translational symmetry of $\ketZ$ and the Hamiltonian, only magnetization peaks at $\bG$ and $\bM$ are allowed. 
When a weak disorder is turned on [Fig.\ref{Fig:FourierAnalysis}(b)], the translational symmetry of the system is broken, and anticipatory peaks appear at all $\textbf{k}$ points, though peaks at $\bG$ and $\bM$ still strongly obscure the others. 
Moreover, in the vicinity of $\bG$ and $\bM$, peaks only appear with a frequency of $\omega=\pm2\omegaS$ or $\omega=\pm\omegaS$ respectively. 
In contrast, {for $\bk$ points far from $\bG$ and $\bM$, the magnetization $M(\textbf{k},\omega)$ peaks at five frequencies $\omega=0,\pm\omegaS,\pm2\omegaS$.}  
As the disorder strength further increases [$\omega=0.5\omegaS$, Fig.~\ref{Fig:FourierAnalysis}~(c)], peaks away from $\bG$ and $\bM$ are almost completely erased by the disorder, and only peaks close to $\bM$ are left with finite residues. 
Finally, if the disorder is strong enough ($W=\omegaS$), all peaks now appear at $\omega=0$ only, as shown in Fig.~\ref{Fig:FourierAnalysis}(d). 

\subsection{Nature of the observed non-ergodic dynamics}

\begin{figure}[t]
\includegraphics[width=\columnwidth]{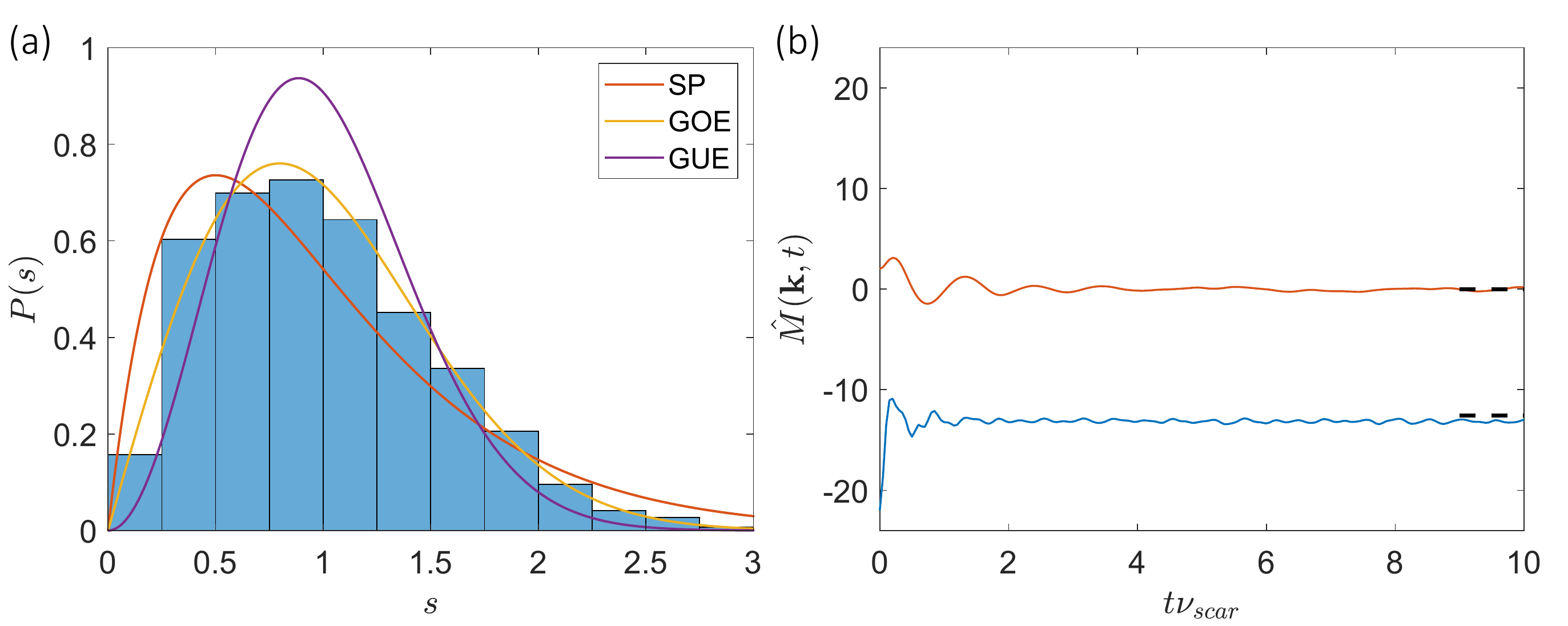}
\caption{\label{Fig:NonInt} (a) The level statistics $P(s)$ of the maximally symmetric sector of the perturbed $6\times 6$ square lattice. 
Here the Hamiltonian is chosen to be $H_\text{clean}$ in Eq.~\eqref{Eq:CleanModel}, so no disorder is present. 
Orange line: special Poisson distribution with $P(s)=4se^{-2s}$. 
Yellow line: GOE with $P(s)=\frac{\pi}2se^{-\pi s^2/4}$. 
Purple line: GUE with $P(s)=\frac{32}{\pi^2}s^2e^{-4s/\pi}$. 
(b) Long-time dynamics of $\hat{M}(\textbf{k},t)$ with $\mathbf k=\bG$ (blue line) and $\bk=\bM$ (orange line), respectively, starting from states with only one excitation.
The canonical (microcanonical) value of $\hat{M}(\boldsymbol{\Gamma},t)$ in (b) is -12.811 (-12.569), while the canonical and the microcanonical values of $\hat{M}(\mathbf{M},t)$ in (b) are exactly 0. We mark the corresponding microcanonical values of $\hat{M}(\textbf{k},t)$ by black dashed lines in (b).
}
\end{figure}

{
We emphasize that the observed nonergodicity and periodic revival in Fig.~\ref{Fig:Magnetization}(a) does not arise from a form of integrability. 
To show this, we first study the level statistics of the maximally symmetric sector of the model in Eq.~\eqref{Eq:CleanModel}, which describes a disorder free system.   
Specifically, we define 
\begin{align}
    s_n = (E_n - E_{n-1})/\expval{s}, 
\end{align}
where $\expval{s}$ is the average level spacing across the entire energy spectrum. 
In a thermal system, the level spacing can follow two typical distributions, depending on whether the time-reversal symmetry (TRS) is present or not. 
In the presence of TRS, the level spacing statistics follows the Gaussian orthogonal ensemble (GOE), with $P(s)=\frac{\pi}2se^{-\pi s^2/4}$. 
In the absence of TRS, however, the level spacing statistics follows the Gaussian unitary ensemble (GUE), with $P(s)=\frac{32}{\pi^2}s^2e^{-4s/\pi}$. 
In our model, the Hilbert space dimension of the maximally symmetric sector in a $6\times4$ lattice is too small (dim$=344$) to carry out a meaningful statistical study, and thus a $6\times 6$ lattice (dim$=9702$) is a more suitable system to clarify this issue. 
As shown in Fig.~\ref{Fig:NonInt}(a), the spectrum of our model perfectly fits GOE, indicating the absence of integrability. 
In fact, the level statistics in the pristine 2D PXP model (i.e., the one described by $H_\text{PXP}$) has been evaluated in Ref.~\cite{Lin2020_PRB}, and it also follows GOE. 
Therefore, we can conclude that the periodic revivals in our model in the clean limit do not arise from integrability. 
}

{
To further demonstrate that the observed revivals arise from scar states, here we demonstrate that quench dynamics starting from the other initial states will result in a fast relaxation. 
In Fig.~\ref{Fig:NonInt}(b) we show the quench dynamics from a state with only one excitation, and we can indeed see that there are no periodic revivals. 
Besides, to benchmark the thermalization process, we also calculated the long-time expectation value of $\hat{M}(\bk,t)$ using the microcanonical ensemble and the canonical ensemble. 
Specifically, the microcanonical ensemble is obtained within the energy window $[E-\Delta E,E+\Delta E]$ where $E$ is the energy expectation of $\ketZ$, $\Delta E$ is $2\%$ of the width of the spectrum, and the inverse temperature $\beta$ for the canonical ensemble is determined by $E=\mathrm{tr}[e^{-\beta}H]/\mathrm{tr}[e^{-\beta}]$. 
By these two means, we can see the expectation in Fig.~\ref{Fig:NonInt}(b) does relax to the equilibrium value, suggesting that the revival of $\ketZ$ does not stem from a form of integrability.}

\section{Possible localization transitions \label{Section:MBL}}

{
Previous discussions show that it takes substantial disorders ($W\sim \omegaS$) to completely destroy the periodic revivals of the 2D PXP model. 
We now study the fate of the 2D PXP model at even stronger disorders. 
In particular, a natural question to ask is whether the PXP model can become many-body localized as the disorder becomes even stronger. 
We will use both the level statistics and EE to study this problem. 

We first look at the level spacing ratio, defined as  
\begin{align}
r_n=\min(\delta E_n,\delta E_{n+1})/\max(\delta E_n,\delta E_{n+1}), 
\end{align}
where $\delta E_n=E_{n+1}-E_n$ is the energy gap between two adjacent energy eigenvalues $E_n$ and $E_{n+1}$. 
In the thermal phase, the spectrum follows the GOE if TRS is present, and the average of level spacing ratio $\overline r$ approaches $\overline r\approx0.53$. 
In contrast, in a thermal system without TRS the spectrum will follow the GUE and $\overline r$ approaches $\overline r\approx 0.60$. 
Finally, we have $\overline r\approx0.39$ in the MBL phase because the spectrum obeys the Poisson distribution. 
In Fig.~\ref{Fig:r_EE}(a), the blue line shows two key features that distinguish the present model from other models having MBL phases (e.g. the XXZ model~\cite{Znidaric2008_PRB}). 
First, $\overline r$ never approaches $0.39$ as $W$ increases. 
Instead, $\overline r$ seems to approach $0.53$ in the $W\to\infty$ limit, implying that the system eventually resides in the thermal phase even in this asymptotic scenario, which is also confirmed by the EE in Fig.~\ref{Fig:r_EE}(c). 
Second, $\overline r$ seems to approach $0.60$ at first and eventually falls back to $0.53$. 
Having shown that the clean PXP model $H_\text{clean}$ in Eq.~\eqref{Eq:CleanModel} follows GOE (see Fig.~\ref{Fig:NonInt}), the blue line in Fig.~\ref{Fig:r_EE}(a) indicates that in the presence of the disorder term $H_w$ the system first loses TRS and regains it later on as $W$ increases. 
In the following two sections, we will explain these two features respectively. 

\subsection{A necessary condition for MBL}

\begin{figure}[t]
\includegraphics[width=\columnwidth]{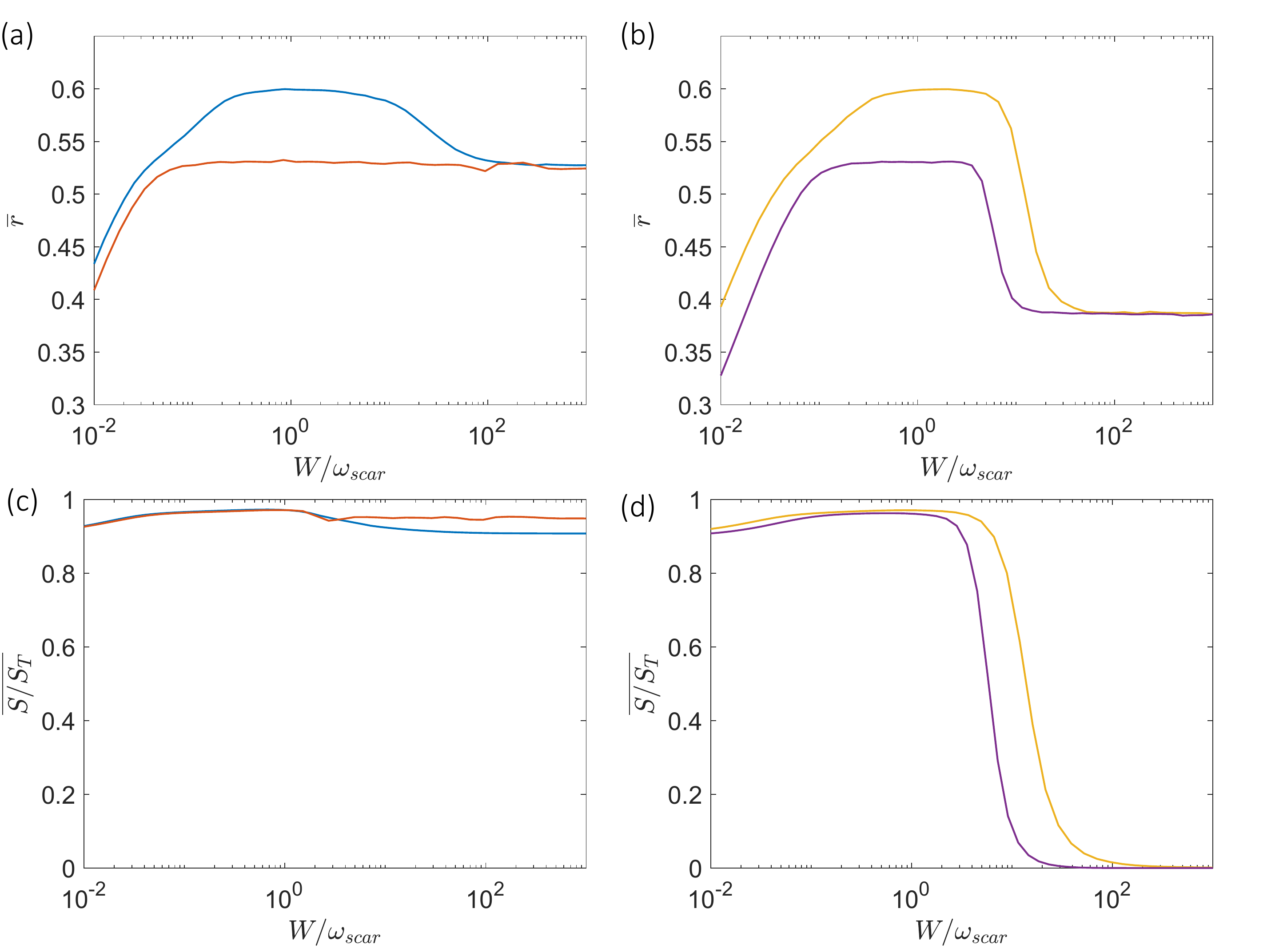}
\caption{\label{Fig:r_EE} (a) and (b) plot the level statistics and (c) and (d) plot the EE. 
In (a) and (c), the blue lines represent the system with disorder $H_w$ and the orange lines represent the system with disorder $H_x$.  
In (b) and (d), the yellow lines represent the system with disorder $H_p$ and the purple lines represent the system with disorder $H_z$.  
All results are obtained in a $6\times4$ square lattice and averaged over 20 disorder realizations.
}
\end{figure}

To explain the absence of MBL phases in the disordered PXP model in Fig.~\ref{Fig:r_EE}(a), we first point out a crucial difference between the disordered PXP model and the disordered Heisenberg XXZ chain, which is known to have an MBL phase. 
Due to the kinematic constraint, the potential operators on different lattice sites $\mathcal P\sigma_{\mathbf r}^a\mathcal P$ in the disordered PXP model may not commute with each other (although the potential operators of different sites in the same sublattice do commute). 
Hence, there exist no common eigenstates for these potential operators, which is different from the disordered Heisenberg XXZ chain. 
As a result, in the PXP model we cannot find a fixed set of basis that can diagonalize the disorder term for any disorder realizations, while in the disordered Heisenberg XXZ chain, we can always find such a basis and the eigenstates do approach the basis asymptotically. 
Therefore, we do not anticipate an MBL phase in our model $\Hfull$ in Eq.~\eqref{Eq:FullModel} even for extremely large $W$ due to this frustration. 

Following the above discussion, we now propose a necessary condition for MBL to occur in a disordered PXP model: 
The potential operators of different sites must commute with each other. 
To support this claim, we replace the disorder term $H_w$ in Eq.~\eqref{Eq:FullModel} with three other types of disorder potentials, 
\begin{equation}
\begin{aligned}
    H_x&=\sum_{\mathbf r}h_{\mathbf r}^x\sigma_{\mathbf r}^x,\quad 
    H_z=\sum_{\mathbf r}h_{\mathbf r}^z\sigma_{\mathbf r}^z,\\
    H_p&=\sum_{\mathbf r \in \mathrm A,a\neq z}h_{\mathbf r}^a\sigma_{\mathbf r}^a+\sum_{\mathbf r \in \mathrm B,a= z}h_{\mathbf r}^a\sigma_{\mathbf r}^a.
\end{aligned}
\end{equation}
For $H_x$, we exclude the effects of $\sigma_y,\sigma_z$ from the full disorder $H_w$ and retain the noncommutativity between different sites. 
For $H_z$, we exclude the effects of $\sigma_x,\sigma_y$ and hence the potential operators naturally commute with each other. 
For $H_p$, we provide another way to gain the commutativity between different sites: 
We turn off $h^z$ on A sublattice and turn off $h^x,h^y$ on B sublattice, and this disorder can be viewed as half of the full disorder. 
As shown in Fig.~\ref{Fig:r_EE}(a), when we choose $H_x$, we can see $\overline r$ reaches $0.53$ for $W>0.1\omegaS$. 
In contrast, we find that in the presence of $H_z$ (instead of $H_w$) $\overline r$ in Fig.~\ref{Fig:r_EE}(b) shows a clear transition from GOE to the Poisson distribution at $W\sim 5\omegaS$. 
Likewise, in the presence of $H_p$ the $\overline{r}$ shows a similar transition from GUE to the Poisson distribution at $W\sim 20\omegaS$.  
The EE results in Figs.~\ref{Fig:r_EE}(c) and~\ref{Fig:r_EE}(d) of the corresponding systems again reveal the same picture.

\subsection{Transition between GOE and GUE}

We now discuss the transition between GOE and GUE observed in Fig.~\ref{Fig:r_EE}(a). 
First, we study the level statistics $P(s)$ of the disordered PXP model $\Hfull$ in Eq.~\eqref{Eq:FullModel} to exclude the possibility of finite size effects. 
In particular, we find that the model $\Hfull$ with $W=2\omegaS$ [see Fig.~\ref{LS}(a)] ($W=200\omegaS$ [see Fig.~\ref{LS}(b)]) fits GUE (GOE) very well, suggesting that the same result is likely to hold in the thermodynamic limit. 
Hence, the system seems to lose TRS for a moderate $W$, but retains TRS for small or large $W$. 

\begin{figure}[t]
\includegraphics[width=\columnwidth]{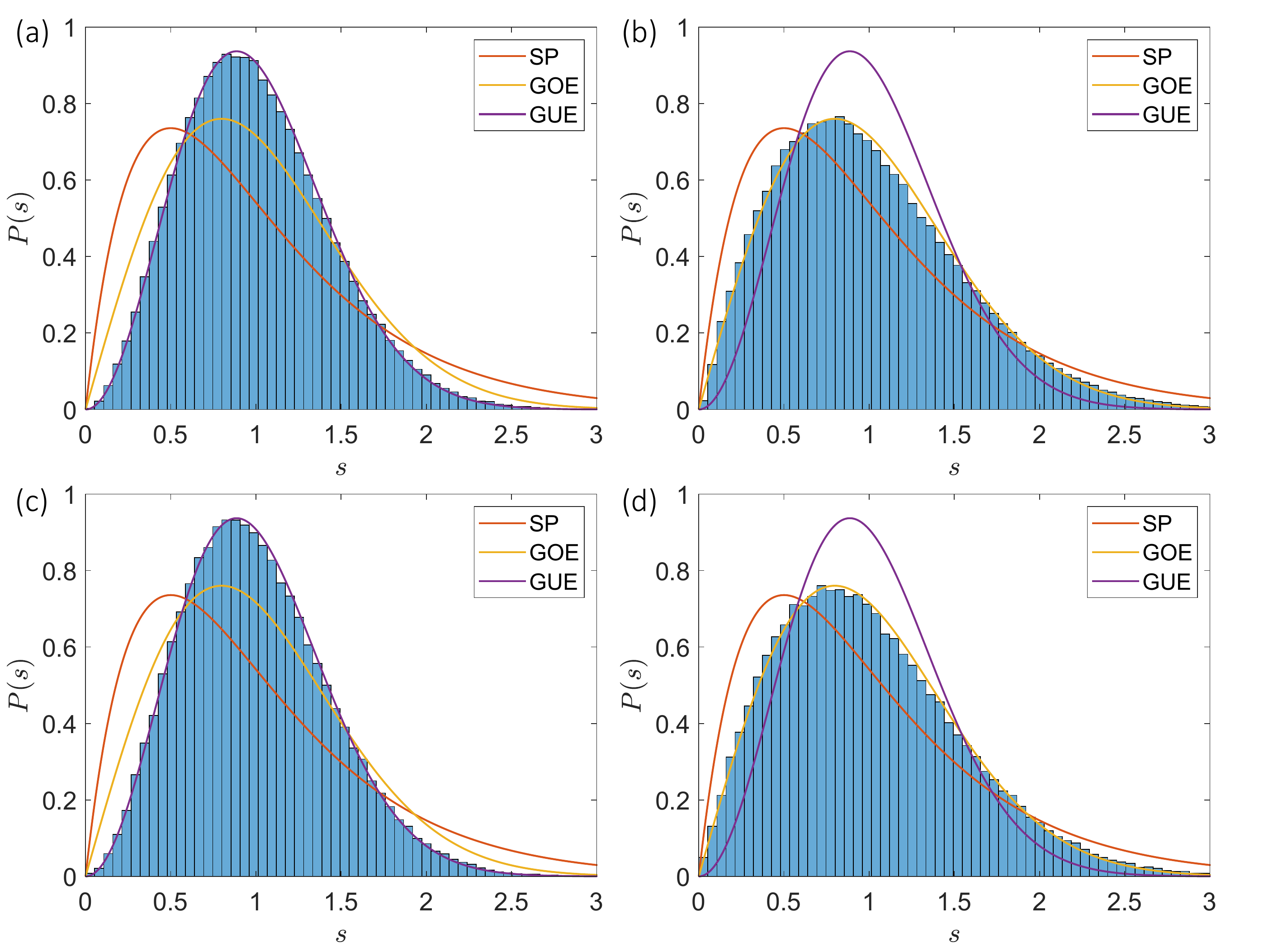}
\caption{\label{LS} (a): $P(s)$ of $\Hfull$ in Eq.~\eqref{Eq:FullModel} with $W=2\omegaS$. 
(b): $P(s)$ of $\Hfull$ in Eq.~\eqref{Eq:FullModel} with $W=200\omegaS$. 
(c): $P(s)$ of $H_1$ with $W=5\omegaS$. 
(d): $P(s)$ of $H_2$ with $W=5\omegaS$. 
All results are obtained in a $6\times4$ square lattice and averaged over $40$ disorder realizations.}
\end{figure}

We speculate that the reason for the presence or absence of TRS is that $\delta H$ (the stabilizer) and $H_w$ (the disorder term) in $\Hfull$ [see Eq.~\eqref{Eq:FullModel}] possess different types of TR symmetries. 
For the disordered pristine PXP model, we have 
\begin{align}
H' 
&\equiv H_\text{PXP} + \mathcal{P}H_w\mathcal{P} \notag\\
&=\mathcal{P}\left[\sum_{\mathbf r}(1+h_{\mathbf r}^x)\sigma_{\mathbf r}^x+h_{\mathbf r}^y\sigma_{\mathbf r}^y+h_{\mathbf r}^z\sigma_{\mathbf r}^z\right]\mathcal{P}. 
\end{align}
Noticing that $[\sigma_{\mathbf r}^z,\mathcal P]=0$, we can rotate each site along the $\hat{z}$ axis. 
That is, there exists a real number $\theta_i$ for each site that satisfies
\begin{equation}
UH' U^{-1}=\mathcal{P}\left[\sum_{\mathbf r}\sqrt{(1+h_{\mathbf r}^x)^2+(h_{\mathbf r}^y)^2}~\sigma_{\mathbf r}^x +h_{\mathbf r}^z\sigma_{\mathbf r}^z\right]\mathcal{P}, \notag
\end{equation}
where $U=e^{-i\sum_{i}\theta_i\sigma_i^z}$. 
Hence, we know for the pristine model we can just consider a real Hamiltonian, or equivalently speaking, the TR symmetry can be captured by the operator $KU$, where $K$ is the complex conjugate, and the specific form of $U$ depends on the disorder realizations. 
In contrast, the extra projector terms in $\delta H$ makes it impossible to incorporate $\delta H$ into the rotation $U$ defined above. 
Moreover, these projection terms leaves $K$ as the only possible TR symmetry for $\delta H$.  
Therefore, if the full model $\Hfull$ is dominated by either $\delta H$ or $H_w$, corresponding to small or large $W$, the system possesses the TR symmetry defined by $K$ or $KU$, respectively, and hence would still follow GOE. 
However, if neither $\delta H$ nor $H_w$ is dominant, the entire system has no TR symmetry and hence would follow GUE. 
The above picture explains the transition between GOE and GUE in the original Hamiltonian $\Hfull$ in Eq.~\eqref{Eq:FullModel}, as well as the $\overline r=0.6$ result in Fig.~\ref{Fig:r_EE}(b) when we replace $H_w$ by $H_p$ in $\Hfull$. 
The above picture also explains why there is no GUE when we turn on $H_x$ only (the TR symmetry is captured by just $K$ in this case) in Fig.~\ref{Fig:r_EE}(a).

\begin{figure*}[!]
    \includegraphics[width=\textwidth]{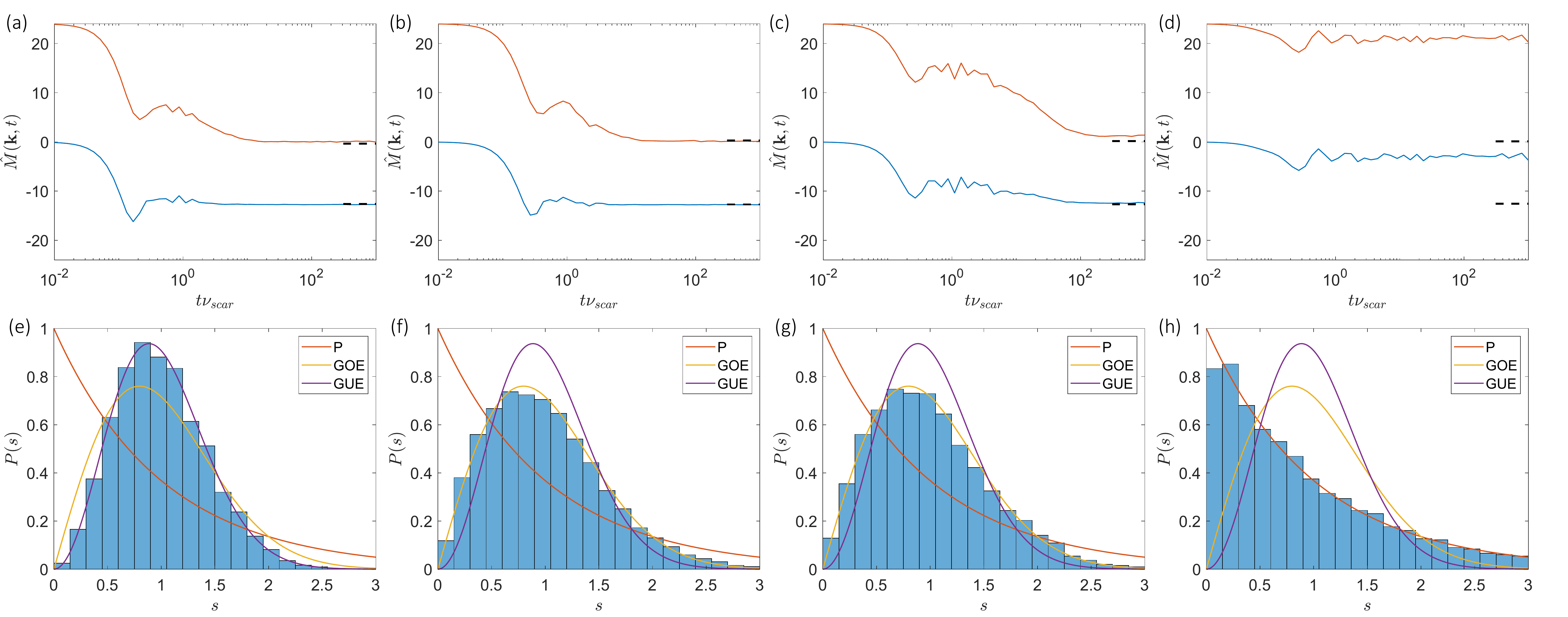}
    \caption{\label{Dyn_LS} Upper panel: the $\hat M(\mathbf k,t)$ of $\ketZ$ at $\mathbf k=\bG$ (the blue lines) and $\mathbf k=\bM$ (the orange lines). 
    Lower panel: The level statistics $P(s)$ of corresponding upper panels. 
    Compared to Fig.~\ref{LS}, the special Poisson distribution is now replaced with the Poisson distribution $P(s)=e^{-s}$. 
    (a): Quench dynamics of $\ketZ$ using $\Hfull$ with $W=2\omegaS$ in $H_w$. 
    (b): Quench dynamics of $\ketZ$ using $H_d$ with $\alpha=0.5$. 
    (c): Quench dynamics of $\ketZ$ using $H_d$ with $\alpha=1$. 
    (d): Quench dynamics of $\ketZ$ using $H_\text{clean} + H_z$ with $W=10\omegaS$. 
    All results are obtained from a fixed disorder realization of $h_{\mathbf r}^a$ in a $6\times4$ square lattice.
    Finally, the canonical (microcanonical) values of $\hat{M}(\boldsymbol{\Gamma},t)$ in (a)-(d) are $-12.766$ ($-12.602$), $-12.770$ ($-12.703$), $-12.694$ ($-12.674$) and $-12.548$ ($-12.578$), respectively. 
    The canonical (microcanonical) values of $\hat{M}(\mathbf{M},t)$ in (a)-(d) are $0.0526$ ($-0.326$), $-0.004$ ($0.271$), $0.001$ ($0.175$) and $0.065$ ($0.121$). 
    We also mark the aforementioned microcanonical values of $\hat{M}(\textbf{k},t)$ by black dashed lines in (a)-(d). 
    }
\end{figure*}

To further validate this idea, we study two special Hamiltonians:
\begin{equation*}
    H_1=\delta H+\mathcal P \left[\sum_{\mathbf r ,a\neq z}h_{\mathbf r}^a\sigma_{\mathbf r}^a\right] \mathcal P, \; 
    H_2=\mathcal P \left[\sum_{\mathbf r ,a\neq z}h_{\mathbf r}^a\sigma_{\mathbf r}^a\right] \mathcal P,
\end{equation*}
with $W=5\omegaS$ in both models, although the specific value of $W$ does not affect the behavior of $H_2$. 
With this choice, we can see the effect of $\delta H$ clearly, though the scale of $\delta H$ is much smaller than the disorder. 
In Figs.~\ref{LS}(c) and~\ref{LS}(d), we find that the level spacing statistics of $H_1$ and $H_2$ perfectly fits GUE and GOE, respectively, just as we expected.

\subsection{Quench dynamics with different disorders}

In the previous sections, we have shown that for an arbitrary $W$ the Hamiltonian $\Hfull$ follows either GOE or GUE, both of which are usually regarded as a signature of a thermal phase. 
However, from a dynamical perspective, the scenario is more complicated in the present model.  

To illustrate this point, we study the dynamics of $\ketZ$ by looking at $\hat M(\mathbf k,t)$ with $\mathbf k=\bG,\bM$.
First, we consider a system governed by $\Hfull$ with $W=2\omegaS$. 
Figure~\ref{LS}(a) suggests that the system resides in the thermal phase (and follows GUE). 
This result is confirmed by the dynamical result in Fig.~\ref{Dyn_LS}(a), as the magnetization of $\ketZ$ relaxes to the equilibrium (microcanonical) value rather quickly. 
Next we consider a system governed by $\Hfull$ with $W = 200 \omegaS$, which, according to Fig.~\ref{LS}(b) fits the GOE.  
In this large $W$ limit, the GOE nature of the system stems from the GOE nature of disorder term $H_w$. 
As a result, in this limit we can effectively study a model with the disorder only, 
\begin{equation}
\begin{aligned}
    H_d=\mathcal{P} \left[\sum_{\mathbf r ,a}h_{\mathbf r}^a\sigma_{\mathbf r}^a\right]\mathcal{P},
\end{aligned}
\end{equation}
where $h_{\mathbf r}^x,h_{\mathbf r}^y$ are uniformly distributed in $[-1,1]$, while $h_{\mathbf r}^z$ is uniformly distributed in $[-\alpha,\alpha]$ and $\alpha$ is the relative strength of the $\hat{z}$ disorder. 
When $\alpha=0.5$, Fig.~\ref{Dyn_LS}(b) shows that the magnetization can still relax to the equilibrium value.  
As we increase $\alpha$ to $\alpha=1$ [see Fig.~\ref{Dyn_LS}(c)], corresponding to the extreme case of $H_w$, the magnetization relaxes much slower than that in Fig.~\ref{Dyn_LS}(b).
Besides, the saturation value of the magnetization deviates from its equilibrium value, indicating that the system is not completely thermal. 
In contrast, the level statistics of all above three cases follows either GOE or GUE, as shown in Figs.~\ref{Dyn_LS}(e)-\ref{Dyn_LS}(g). 
Finally, we study the dynamical feature of $\ketZ$ in the MBL phase for comparison. 
Specifically, we consider a system governed by $H_\text{clean} + H_z$ with $W=10\omegaS$. 
The level statistics result in Fig.~\ref{Dyn_LS}(h) obeys Poisson distribution and the magnetization in Fig.~\ref{Dyn_LS}(d) is clearly localized. 
The above results show that, the behavior of $H_d$ with $\alpha=1$, or similarly, $\Hfull$ with $W>200\omegaS$, is distinct from either the thermal phase or the MBL phase, although its spectrum fits GOE well. 
Moreover, as the $\hat{z}$ disorder becomes more dominant, i.e. increasing $\alpha$, the dynamics of $\ketZ$ becomes more nonergodic. 
We remind the readers that the results shown in this section are obtained from a $6\times4$ lattice, which is a finite system. 
We cannot exclude the possibility that the apparent disagreement between the spectral [Fig.~\ref{Dyn_LS}(c)] and dynamical properties [Fig.~\ref{Dyn_LS}(g)] may vanish in the thermodynamic limit. 
}

\section{Summary and outlook}
In this work we studied the fate of quantum many-body scar states in the 2D PXP model against random disorders. 
In particular, we showed that the scar states in a 2D square lattice are stable against random disorders up to $W\sim 0.1\omegaS$, before eventually being erased. 
We demonstrated {the different regimes} of the PXP model by studying the magnetization of the system $M(\br,t)$. 
We also studied whether the 2D PXP model enters a many-body localized phase at strong disorders. 
We found that the behavior of the PXP model strongly depends on the type of disorder present in the system. 
Finally, we have also studied the results in a 2D honeycomb lattice, and found qualitatively similar results (see Appendix~\ref{Appendix}). 

Our work also raises several interesting questions for future studies. 
For example, it is interesting to understand the stability of scar states in the presence of quasiperiodic potentials instead of random disorders. 
In addition, it is not clear whether the intermediate phase found in Fig.~\ref{Fig:r_EE} has some similarity to the intermediate phase identified during the transition between a thermal and an MBL phase in a quasiperiodic system~\cite{Hsu2018_PRL,Kohlert2019_PRL,Xu2019_PRR}. 
Future experiments may help us verify the results in our work and also help us understand these open questions.

\section{Acknowledgements}
X.L. thanks Dong~E.~Liu for fruitful discussions. 
X.L. acknowledges support from City University of Hong Kong (Project No.~9610428), the National Natural Science Foundation of China (Grant No.~11904305), as well as the Research Grants Council of Hong Kong (Grant No.~CityU~21304720). 
Y.W. acknowledges the support from the National Natural Science Foundation of China under Grant No.~11874292, No.~11729402, and No.~11574238.


\appendix
\section{Scar states in a 2D honeycomb lattice \label{Appendix}}

In this appendix, we study the stability of quantum many-body scars against random disorders in a 2D honeycomb lattice. 
In particular, the PXP Hamiltonian $H_{\mathrm{PXP}}$ and the disorder potential defined in the main text are still applicable to the honeycomb lattice, which we quote below: 
\begin{align}
    H_{\mathrm{PXP}}=\frac{\Omega}{2}\sum_{\br}\sigma_{\br}^x\prod_{\langle\br',\br\rangle}P_{\br'}^-, \quad 
    H_w = \sum_{\br}h_a(\br)\sigma_{\br}^a. 
\end{align}
However, the stabilizer $\delta H$ defined in the main text no longer applies to the honeycomb lattice. 
Instead, there exists a different perturbation $\delta H$ to enhance the weak nonergodicity in the honeycomb lattice~\cite{Michailidis2020_PRR},
\begin{equation}
\begin{aligned}
    \delta H=\sum_{\br}\sigma_{\br}^x\left[aP_{\br}^l+bP_{\br}^2\right]\prod_{\langle\br',\br\rangle}P_{\br'}^-,
\end{aligned}
\end{equation}
where the projection operators are given by 
\begin{equation}
\begin{aligned}
    P_{\br}^l&=\sum_{\langle\langle\br',\br\rangle\rangle}P_{\br'}^-, \quad 
    P_{\br}^2&=\sum_{\langle\br',\br\rangle}P_{\br'}^-\qty\bigg[\prod_{\substack{\langle\br'',\br'\rangle,\\\mathbf r''\neq\mathbf r}}P_{\br''}^-].
\end{aligned}
\end{equation}
In the above equation, ${\langle\langle\br',\br\rangle\rangle}$ indicates summing over all next nearest neighbors of site $\br$. 
In addition, $a$ and $b$ are optimized according to the revival of the fidelity 
$\abs{\ip{\mathbb{Z}_2}{\mathbb{Z}_2(t)}}$~\cite{Michailidis2020_PRR}. 
Note that there are two sublattices in the honeycomb lattice, which we name as the $A$ and $B$ sublattice, respectively. 
Hence $\ketZ$ in the honeycomb lattice is defined as the state with all atoms on one sublattice getting excited while those on the other sublattice remaining in the ground state. 
In this appendix we specifically choose $\ketZ$ to be the state with all atoms on the $A$ sublattice being excited. 
In a $3\times4$ honeycomb lattice, we find that $a=0.0160\Omega$ and $b=0.0321\Omega$, which is similar to the results in a $3\times3$ and a $4\times4$ honeycomb lattice, affirming the previous results~\cite{Michailidis2020_PRR} that these optimized parameters are independent of not only the size but also the shape of the honeycomb lattice. 

\begin{figure}[t]
\includegraphics[width=\columnwidth]{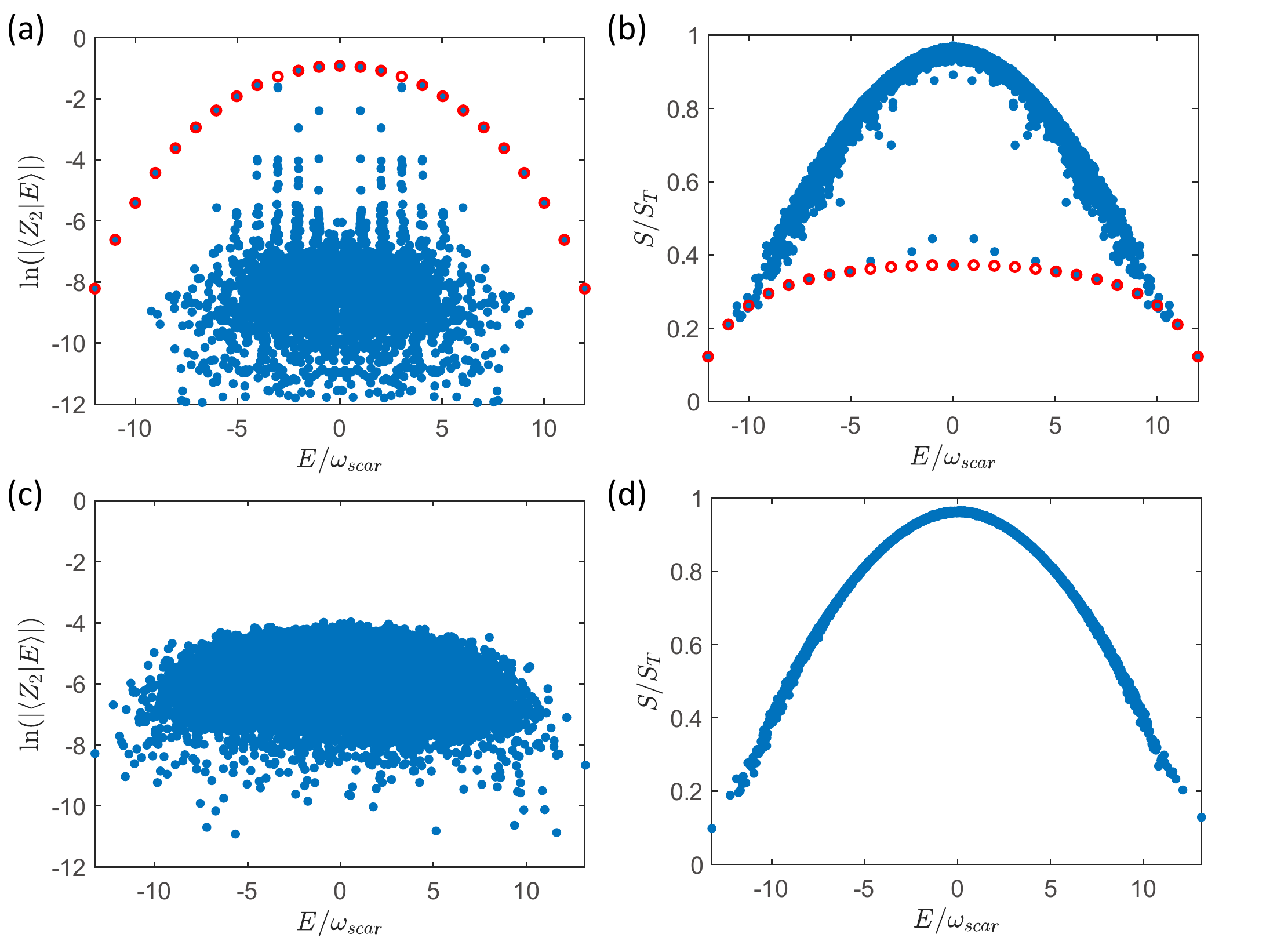}
\caption{\label{Fig:Honeycomb} (a) The overlap between the eigenstates of the PXP model and the $\ketZ$ state in a $3\times4$ honeycomb lattice. 
(b) The EE of all eigenstates in the same model. 
(c) and (d) are the respective quantities in a system with a fixed disorder realization with $W=0.5\omegaS$. 
Blue dots are obtained by exact diagonalization methods and red circles in (a) and (b) are results from the FSA approximation.}
\end{figure}

In our calculations we find that the honeycomb lattice exhibits qualitatively similar properties to the square lattice in terms of the overlap with $\ketZ$ and EE. 
In particular, in Fig.~\ref{Fig:Honeycomb} we observed a similar structure for the scar states in a clean system and verified that it is eventually erased by strong disorders.

\begin{figure*}[!]
\includegraphics[width=2\columnwidth]{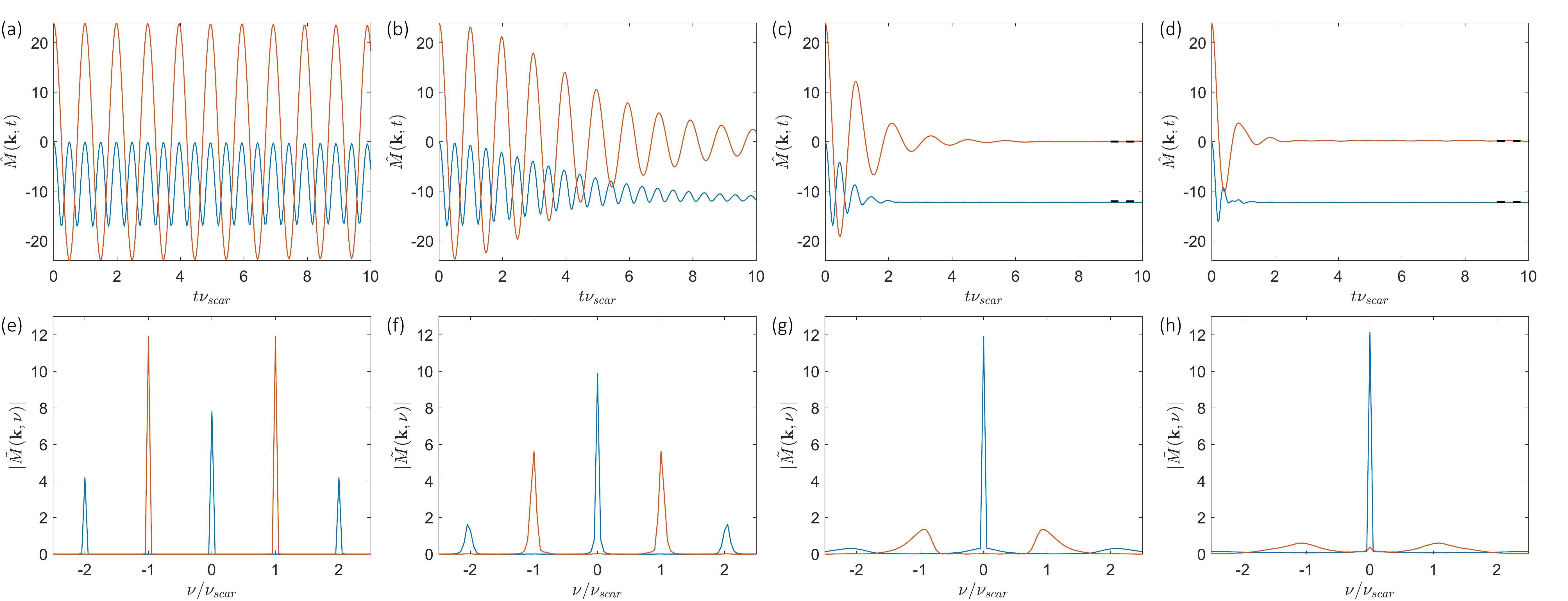}
\caption{\label{Fig:Magnetization_HC} (a)-(d) are the results in a $3\times4$ honeycomb lattice with a fixed disorder realization. 
The disorder strength for each panel [from (a) to (d)] is $W=0,0.1\omegaS,0.5\omegaS,\omegaS$, respectively. 
The lower panels show the Fourier transformation of the corresponding upper panels. 
The blue lines represent $\hat{M}_+$ or $\tilde{M}_+$, and the orange lines represent $\hat{M}_-$ or $\tilde{M}_-$, respectively. 
The canonical (microcanonical) values of $\hat{M}_+$ in (c) and (d) are -12.315 (-12.047) and -12.329 (-12.105).
The canonical (microcanonical) values of $\hat{M}_-$in (c) and (d) are 0.064 (0.0332) and 0.203 (0.166). 
We mark the corresponding microcanonical values of $\hat{M}_{\pm}$ by black dashed lines in (c) and (d). 
}
\end{figure*}

\begin{figure*}[!]
\includegraphics[width=2\columnwidth]{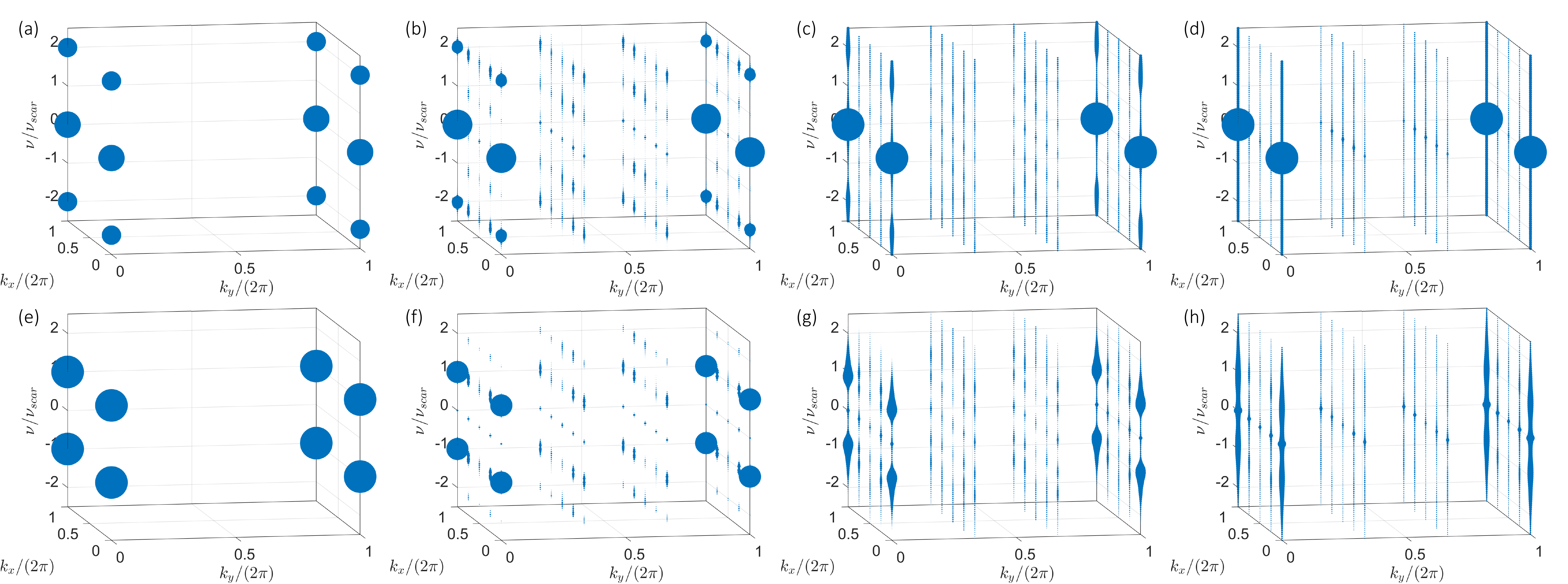}
\caption{\label{Fig:FourierAnalysis_HC} (a)-(d) plot $\tilde M_+$ in a $3\times4$ honeycomb lattice with $W=0,\,0.1\omegaS,\,0.5\omegaS,\,\omegaS$ respectively. 
All results are averaged over $20$ disorder realizations. 
The radii of the spheres are proportional to $\abs*{\tilde M(\textbf{k},\omega)}^{1/2}$. 
The lower panels display the corresponding $\tilde M_-$ for each upper panel. }
\end{figure*}

We now discuss the dynamics of the magnetization $M(\br,t)=\langle \mathbb{Z}_2(t)|\sigma_{\br}^z|\mathbb{Z}_2(t)\rangle$ in this $3\times 4$ honeycomb lattice.
Since there are two sublattices in the honeycomb lattice, we can define two independent Fourier transformations, one for each sublattice,
\begin{equation}
\begin{aligned}
    \hat M_j(\textbf{k},t)&=\sum_{\br\in j}e^{i\textbf{k}\cdot\br} M(\br,t),\\
    \tilde M_j(\textbf{k},\omega)&=\int_{-\infty}^{+\infty}M_j(\textbf{k},t)e^{-i\omega t}\mbox dt.
\end{aligned}
\end{equation}
In the above equation $j=\mathrm{A},\mathrm{B}$ represents the two sublattices. 
In addition, we have $\textbf{k}=k_x \textbf{b}_1+k_y \textbf{b}_2$, where $\textbf{b}_1,\textbf{b}_2$ are the bases of the reciprocal space. 
Further, we introduce the following combinations
\begin{equation}
\begin{aligned}
    \hat M_{\pm}(\textbf{k},t)&=\hat M_{\mathrm{A}}(\textbf{k},t)\pm \hat M_{\mathrm{B}}(\textbf{k},t),\\
    \tilde M_{\pm}(\textbf{k},\omega)&=\tilde M_{\mathrm{A}}(\textbf{k},\omega)\pm \tilde M_{\mathrm{B}}(\textbf{k},\omega).
\end{aligned}
\end{equation}
It turns out that $\hat M_{\pm}(\textbf{0},t)$ corresponds to $\hat M(\textbf{k},t)$ with $\textbf{k}=(0,0)$ and  $\textbf{k}=(\pi,\pi)$ in the square lattice, respectively. 
We find from Fig.~\ref{Fig:Magnetization_HC} and Fig.~\ref{Fig:FourierAnalysis_HC} that as the disorder strength increases, there is also a transition from weak ergodicity breaking to a thermal phase.
In particular, the strongest peak only appears at $\textbf{k}=\textbf{0}$, and $\tilde M_{\pm}(\textbf{0},\omega)$ peaks at $\omega=\pm2\omegaS$ or $\omega=\pm\omegaS$ initially, while as disorder turned on, all peaks move to $\omega=0$.

\begin{figure}[!]
\includegraphics[width=\columnwidth]{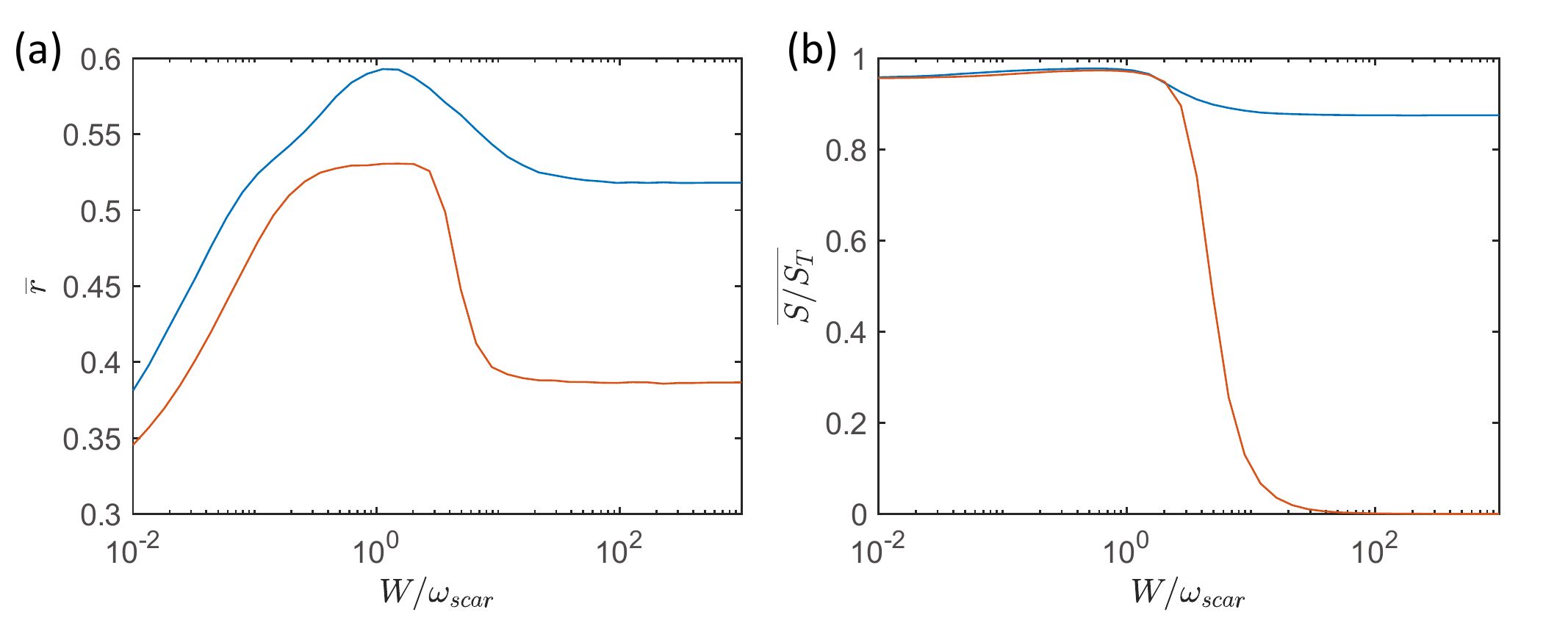}
\caption{\label{Fig:MBL_HC} (a) and (b) plot the EE and level statistics in a $3\times3$ honeycomb lattice. 
The data is averaged over 400 disorder realizations. 
The blue lines represent the results with $h^x,\,h^y,\,h^z$ distributed uniformly in $[-W/2,W/2]$, while the orange lines represent the results with $h^x=h^y=0$ and $h^z$ distributed uniformly in $[-W/2,W/2]$.}
\end{figure}

Finally, we discuss the behavior of the system at very strong disorders. 
In Fig.~\ref{Fig:MBL_HC}, we plot the EE and level statistics in a $3\times3$ honeycomb lattice. 
For the EE, the lattice is divided into a left part (a $2\times3$ honeycomb lattice) and a right part (a $1\times3$ honeycomb lattice) with dimension $D_L=260$ and $D_R=18$, respectively, and therefore the EE in thermal phase is given by $S_T=\ln D_L-D_R/(2D_L)$~\cite{Khemani2017_PRL}. 
It turns out that the result again depends on the type of disorder we introduce. 
In particular, if $h^x$, $h^y$, and $h^z$ are all present, a strong disorder still cannot localize the system. 
In contrast, if $h^x=h^y=0$ and only $h^z$ is present, the small system seems to approach an MBL phase quickly. 
We believe that the different behaviors of different disorder potentials observed here arise from a similar mechanism as the one we discussed in the main text for the square lattice. 
Thus, we do not elaborate further on this topic. 


\bibliographystyle{apsrev4-2}
\bibliography{Scars.bib}



%

\end{document}